\documentclass[preprint,12pt]{elsarticle}
\pdfoutput=1

\usepackage{amssymb}

\makeatletter

\newcommand{\Rmnum}[1]{\expandafter\@slowromancap\romannumeral #1@}
\makeatother
\journal{Nuclear Physics B}
\usepackage{amsmath}
\usepackage{booktabs}
\usepackage[misc]{ifsym}
\usepackage{tikz}
\usepackage{bm}
\tikzstyle{block} = [draw,rectangle,thick,minimum height=2em,minimum width=2em,anchor=west]
\renewcommand{\vec}[1]{\ensuremath{\boldsymbol{#1}}} 
\usepackage{pgfplots}
\begin{document}

\begin{frontmatter}



\title{A special D4-D2-branes configuration on K3 fiberation Calabi-Yau threefolds}


\author[mymainaddress]{Fei Wang\corref{mycorrespondingauthor}}
\cortext[mycorrespondingauthor]{Department of Physics and Mechanics, Qingdao City University}
\ead{fei.wang@qdc.edu.cn}

\address[mymainaddress]{Department of Physics and Mechanics, Qingdao City University, Qingdao, Shandong 266000, China}

\begin{abstract}
	
We construct a special D4-D2-brane configuration on the K3 fibered  Calabi-Yau threefolds: canonical D4-branes on K3 fiberations and another D2-branes on its base $\mathbb{P}_1$, by using the ordinary nonlinear sigma model in two dimensions. We would derive the exact expression of $Ext$ groups of this construction. By using the Gepner model, we achieve the feasibility of the existence of such a configuration and  get more available sheaf's data of this brane system.

\end{abstract}


\end{frontmatter}


\section{introduction}

With the development of mirror symmetry and derived categories, some mathematical concepts deriving from physics have been introduced into the explanation of related physical theories. Coherent sheaves in algebraic geometry, appearing for the physical significance, is described as massless boundary Ramond sector states of D-branes wrapped on complex submanifolds of Calabi-Yau's, with holomorphic gauge bundles. The number of these states should be counted by mathematical objects known as $Ext$ groups \cite{2003D,1998On,2002D}. The nonlinear sigma model have an important development in string, which describes the maps from Riemann surfaces to a Calabi-Yau manifold \cite{2000MIRROR} Therefore, we start with type \Rmnum{2} string on $X_d \times \mathbb{R}^{D,1}$, where $D=9-2d$ and $X_d$ is a Calabi-Yau manifold of complex dimension $d$. The nonlinear sigma model encodes the A-model branes and B-model branes on a Calabi-Yau threefold $X$ where A side is described as the Largrange submanifolds of $X$ and B side is described as the holomorphic complex submanifolds of $X$. B-branes wrapped on the holomorphic submanifolds are D4-branes or D2-branes and B-branes wrapped on the entire Calabi-Yau threefold are D6-branes.  In this paper, we study a special D4-D2-brane system configuration on the K3 fibered Calabi-Yau threefold with D4-branes wrapping on the K3-fiberation and D2-branes wrapping on $\mathbb{P}_1$ base, where $\mathbb{P}_1$ base could also be regarded as fiberation. The well-known BCFT methods compute the massless Ramond sector sepctrum of open strings contracting such D-branes, and then relating that sepctrum to Ext groups. Those spectral sequences are realized directly in BRST cohomology, and so the massless Ramond sector spectrum is counted directly by $Ext$ groups \cite{2002D}. D-branes wrapped on the complex submanifold $i: S \hookrightarrow X$ and $j: T \hookrightarrow X$ of a Calabi-Yau $X$, with holomorphic gauge bundles in $ \mathcal{E}\otimes\sqrt{K^{\vee}_{S}}, \mathcal{F}\otimes\sqrt{K^{\vee}_{T}}$ \renewcommand{\thefootnote}{\roman{footnote}}\footnote{In order to cancel the Freed-Witten anomaly, the sheaves $i_{*}\mathcal{E}$ really corresponds to the Chan-Paton sheaves $ \mathcal{E}\otimes\sqrt{K^{\vee}_{S}}$. Since the Chan-Paton factors couple to the bundle $ \mathcal{E}\otimes\sqrt{K^{\vee}_{S}}$ \cite{1999Anomalies}} (Chan-Paton sheaves), respectively, then massless boundary Ramond sector states should be counted with elements of either $$ Ext^{*}_{X}(i_{*}\mathcal{E}, j_{*}\mathcal{F})$$ or $$ Ext^{*}_{X}( j_{*}\mathcal{F}, i_{*}\mathcal{E}).$$

We cannot count all the $Ext$ groups in the ordinary sigma model on Calabi-Yau threefold at large radius, since it is impossible to count all the BRST cohomology on every generic point. Gepner model \cite{1988Space} appears with special rational points in the moduli space where the boundary conformal fields connecting to the open string is exactly solvable. Gepner model construction at these special points continuously connects to the "large radius". In Sect.2, we use this special sigma model to get more available information of our configuration.

We will consider two K3 fiberation models $6^{2}2^{3}{/}\mathbb{Z}_{8}$ and $10^{2}4^{2}0 {/} \mathbb{Z}_{12}$. By the above algorithm through the modulo orbit groups, the first Calabi-Yau manifold which is defined classically by resolving the singularities of a quasi-homogeneous polynomial hypersurface of total degree 8 in a four-dimensional projective space weighted by $(1,1,2,2,2)$, hereafter denoted $\mathbb{P}(1,1,2,2,2)[8]$. The second one contains the geometric data for a degree 12 projective variety, denoted as $\mathbb{P}(1,1,2,2,6)[12]$. Their quantum geometry has been extensively investigated using mirror symmetry \cite{1993Mirror}.

\section{D4-D2 boundary states construction on K3 fibered Calabi-Yau threefolds}
\subsection{The gauge bundles on D-branes of the B-type model}
The B-type model is a twist of the ordinary nonlinear sigma model in two dimensions, with bulk lagrangian:

$$g_{\mu\nu}\partial\phi^{\mu}\overline{\partial}\phi^{\nu} + ig_{i\overline{j}}\psi^{\overline{j}}_{-}D_{z}\psi^{i}_{-} + ig_{i\overline{j}} + ig_{i\overline{j}}\psi^{\overline{j}}_{+}D_{\overline{z}}\psi^{i}_{+} + R_{i\overline{i}j\overline{j}}\psi^{i}_{+}\psi^{\overline{i}}_{+}\psi^{j}_{-}\psi^{\overline{j}}_{-}. $$
In the B-type model, the worldsheet fermions are twisted to be sections of different bundles \cite{2002D,2000MIRROR,alma9914090123903768}:$$\psi^{\overline{i}}_{\pm} \in \Gamma(\phi^{*}T^{0,1}X)$$ $$\psi^{i}_{+} \in \Gamma(K \times \phi^{*}T^{1,0}X)$$ $$\psi^{i}_{-} \in \Gamma(\overline{K} \otimes \phi^{*}T^{1,0}X),$$ where $K$ is the canonical bundle and $\overline{K}$ is its complex conjugate.
Following \cite{2002D}, we have $$\eta^{\overline{i}} = \psi^{\overline{i}}_{+} + \psi^{\overline{i}}_{-}$$ 
$$\theta_{i} = g_{i\overline{j}}(\psi^{\overline{j}}_{+} + \psi^{\overline{j}}_{-})$$ 
$$\rho^{i}_{z} = \psi^{i}_{+}$$ 
$$\rho^{i}_{\overline{z}} = \psi^{i}_{-}.$$
The boundary conditions on the worldsheet fermions can be expressed as $\eta = 0$ for Dirichlet directions and $\theta_{i} = F_{i\overline{j}}\eta^{\overline{j}}$ for Neumann directions. In the B-type boundary conditions in zero curvature, $TX|_{S}$ splits holomorphically as $TS \oplus \mathcal{N}_{S/X}$, and we have an extension:$$0 \longrightarrow TS \longrightarrow TX|_{S} \longrightarrow \mathcal{N}_{S/X} \longrightarrow 0.$$

The BRST-closed states are of the form
\begin{equation}\label{5}
	b^{j_{1}\cdots j_{m}}_{\overline{i}\cdots \overline{i}_{n}}(\phi)\eta^{\overline{i}_{1}}\cdots\eta^{\overline{i}_{n}}\theta_{j_{1}}\cdots\theta_{j_{m}},
\end{equation}
In order to make mathematical techniques (holomporphic gauge bundles or coherent sheaves) available, we translate these physical operators into mathematical representation: $\eta$ $\in$ antiholomorphic 1-form $\Omega_{\overline{z}}^1$ and $\theta$ $\in$ holomorphic tangent bundle $T_{z}$. We can donate them as$$\eta^{\overline{i}} \thicksim d\overline{z}^{\overline{i}}, \qquad \theta_{i} \thicksim \frac{\partial}{\partial z^{i}}.$$

Thus, the states ($\ref{5}$) are in one-to-one correspondence with bundle-valued differential form$$b^{j_{1}\cdots j_{m}}_{\overline{i}\cdots \overline{i}_{n}}d\overline{z}^{\overline{i}_{1}}\wedge\cdots\wedge d\overline{z}^{\overline{i}_n}\wedge\frac{\partial}{\partial z^{j_{1}}}\wedge\cdots\wedge\frac{\partial}{\partial z^{j_{m}}}.$$ The B-type model bulk states are counted by the sheaf cohomology groups $$H^{n}(X, \wedge^{m}T^{1,0}X).$$

According  to our previous claims the physical relevance of $Ext$ groups, we assume that there is a gaugle bundle $\mathcal{E}$ on one D-brane and another guage bundle on the other D-brane is $\mathcal{F}$, and then the open string spectrum must be counted by the Ext groups \cite{2002D} $$Ext^{*}_{X}(i_{*}\mathcal{E}, i_{*}\mathcal{F}),$$ where $i$ is an inclusion as stated before.

More generally, following the same methods as for the closed string, the local operator of the boundary R-sector states are of the form $$ b^{\alpha\beta j_{1}\cdots j_{m}}_{\overline{i}\cdots \overline{i}_{n}}(\phi)\eta^{\overline{i}_{1}}\cdots\eta^{\overline{i}_{n}}\theta_{j_{1}}\cdots\theta_{j_{m}},$$ where the $\alpha$, $\beta$ are Chan-Paton indices, coupling to the bundles on either D-brane. We have the same dictionary as for the bulk states, with the modifications mentioned above that $\eta$ is the antihomlomorphic 1-forms on submanifold $S$, $\theta$ couple to $\mathcal{N}_{S/X}$, and the $\phi$ zero modes are restrited to $S$. The states above are classified by the sheaf cohomology groups 
\begin{equation}\label{6}
	H^{n}(S, \mathcal{E}^{\vee} \otimes \mathcal{F} \otimes \wedge^{m} \mathcal{N}_{S/X}).
\end{equation}
\subsection{The general conditions of being Calabi-Yau}
Let $\pi : X \rightarrow Y$ be an algebraic fiber space. We assume the base $Y$ of the fiberation is smooth and the fibers are generically smooth and $X$ is also smooth. Let $K_{X}, K_{Y}$ denote the canonical bundles. The relative canonical bundle is $K_{X|Y}$, defined as
\begin{equation}\label{1}
	K_{X|Y} = \pi^{*}K^{-1}_{Y} \otimes K_{X}
\end{equation}
\begin{equation}\label{2}
	K_{X} = \mathcal{O}_{X} \Rightarrow \pi_{*}K_{X|Y} = K^{-1}_{Y}
\end{equation}
\begin{equation}\label{3}
	K_{X|\Sigma} = K_{\Sigma} - c_{1}(N_{X}\Sigma)
\end{equation}
\begin{equation}\label{4}
	c_{1}(N_{X}\Sigma) = K_{\Sigma}
\end{equation}
where $\Sigma$ is the subvariety of X and $c_{1}$ is the first Chern class.

Assume $X$ is as $K3$-fiberation Calabi-Yau threefold, the base $Y$ is necessarily $\mathbb{P}_{1}$ \cite{1999K3}. The equation ($\ref{4}$) becomes in this case of $c_{1}(N_{X}\Sigma) = -2$.

\subsection{Construction of the D4-D2 boundary states on $K3$ fibered Calabi-Yau threefolds}

Assume $S$ is the $K3$ fiberation space and let $i, \pi$: $S \stackrel{i}{\hookrightarrow} X \stackrel{\pi}{\rightarrow} Y$, where $i$ is an including and $\pi$ is a surjection and $Y$ is the $\mathbb{P}_{1}$ base (or the $\mathbb{P}_1$ fiberation space). The configuration, consistent with such a structure of Calabi-Yau varieties, could be that the D4-branes wrap on the $K3$ fiberation space $S$ and the D2-branes wrap on $\mathbb{P}_1$ fiberation space $Y$.

Here, we use the mathematical concept $\emph{sheaf}$\renewcommand{\thefootnote}{\roman{footnote}}\footnote{One motivation for such a extended mathematical concept is that it allows us to convert physics questions into mathematical computation by using more advantageous and advanced techniques.} to model D-branes on large-radius Calabi-Yau manifolds \cite{1999D}, and coherent sheaf (a extended concept of holomorphic gauge bundle) can be used to calculate physical properties, for example open string spectra between D-branes could be counted by Ext groups.

Now, we deduce the classification of these D4-D2 boundary states by sheaf cohomology groups. For the D4-branes wrapped on the K3 fiberation, it is the same as ($\ref{6}$)$$H^{n}(S, \mathcal{E}^{\vee} \otimes \mathcal{F} \otimes \wedge^{m} \mathcal{N}_{S/X}).$$ 

For the D2-branes wrapping on the $\mathbb{P}_{1}$, we assume $\xi, \zeta$ are the gauge bundles on two D2-branes respectively. For the maps $\pi$ of K3 fibered Calabi-Yau threefolds in $S \stackrel{i}{\hookrightarrow} X \stackrel{\pi}{\rightarrow} Y(\mathbb{P}_{1})$, we have the sheaf cohomology groups: $$H^{n}(Y, \xi^{\vee} \otimes \zeta \otimes \wedge^{m} \mathcal{N}_{Y/X}).$$ The cohomology groups of gauge bundles mapping between the D2-branes and the D4-branes could be denoted as $H^{n}(X, Hom(\mathcal{E}, \xi)) \cong H^{n}(X, \mathcal{E}^{\vee} \otimes \xi)$ or $H^{n}(X, Hom(\xi, \mathcal{E})) \cong H^{n}(X, \xi^{\vee} \otimes \mathcal{E})$. So the vertex operator could be denoted as $\xi^{\vee} \otimes \mathcal{E} \otimes \wedge^{m} \mathcal{N}_{S/X}$ on $S$ and $\mathcal{E}^{\vee} \otimes \xi \otimes \wedge^{m} \mathcal{N}_{Y/X}$ on $Y$. Now, we could get the interaction cohomology groups of D4-D2-brane configuration as$$H^{n}(S, \xi^{\vee} \otimes \mathcal{E} \otimes \wedge^{m} \mathcal{N}_{S/X})$$ $$H^{n}(Y, \mathcal{E}^{\vee} \otimes \xi \otimes \wedge^{m} \mathcal{N}_{Y/X}).$$

So far, we use the cohomology groups to count the classification of open string spectrum but not the $Ext$ groups. We can get the relation between them with the curvature vanishing on the gauge bundle and $TX|_{S}$ and $TY|_{Y}$ splitting holomorphically as $TS \oplus \mathcal{N}_{S/X}$ and $TY \oplus \mathcal{N}_{Y/X}$ respectively. It implies that $$ Ext^{n}_{X}(i_{*}\mathcal{E}, i_{*}\mathcal{F}) \cong \sum_{p+q=n} H^{p}(S, \mathcal{E}^{\vee} \otimes \mathcal{F} \otimes \wedge^{q} \mathcal{N}_{S/X})$$  
$$ Ext^{n}_{X}(\pi^{*}\xi, \pi^{*}\zeta) \cong \sum_{p+q=n} H^{p}(Y, \xi^{\vee} \otimes \zeta \otimes \wedge^{q} \mathcal{N}_{Y/X})$$

\begin{equation}\label{7}
	Ext^{n}_{X}(\pi^{*}\xi, i_{*}\mathcal{E}) \cong \sum_{p+q=n} H^{p}(S, \xi^{\vee} \otimes \mathcal{E} \otimes \wedge^{q} \mathcal{N}_{S/X})
\end{equation}

\begin{equation}\label{8}
	Ext^{n}_{X}(i_{*}\mathcal{E}, \pi^{*}\xi) \cong \sum_{p+q=n} H^{p}(Y, \mathcal{E}^{\vee} \otimes \xi \otimes \wedge^{q} \mathcal{N}_{Y/X}),
\end{equation}
where the $Ext$ groups ($\ref{7}$) on $S$ and ($\ref{8}$) on $Y$ classify the D4-D2 branes interaction terms. 

The $Ext$ groups of these two D4-D2-branes interaction terms describe the vertex operators of the boundary R-sector states in BCFT. The moduli space of the target space of the nonlinear sigma model contain no singular points so that it is impossible to count the classes of the bundles on all points. It is impossible for us to obtain concrete representations of these $Ext$ groups, but we can get the expression of these $Ext$ groups in the case of single D-branes.

More generally, we can promote these group representations to other multi-brane interaction systems $(0,2,4,6)$ on non-homogeneous varieties on 6d Calabi-Yau manifold. The direct method to describe these $Ext$ groups is the category of quiver representations. A quiver $Q$ is a directed graph containing a set of vertices $Q_0$, a set of arrow $Q_1$ and maps $h$, $t$: $Q_0 \longrightarrow Q_1$ which specify the head and tail \cite{0Representation,0Representation1,0Representation2}. For example, the multi-branes interaction systems for our configuration could be described as

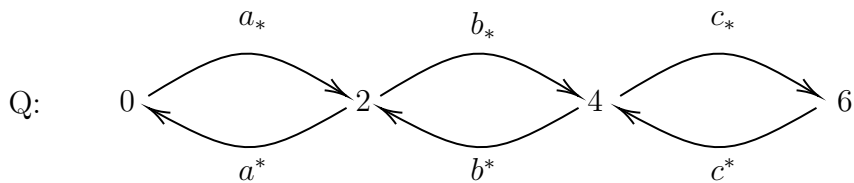
\tikzset{every picture/.style={line width=0.75pt}} 
\begin{center}
\begin{tikzpicture}[x=0.75pt,y=0.75pt,yscale=-1,xscale=1]
	
	\draw    (144,82) .. controls (189.87,53.78) and (197.19,53.5) .. (242.61,81.15) ;
	\draw [shift={(244,82)}, rotate = 211.29] [color={rgb, 255:red, 0; green, 0; blue, 0 }  ][line width=0.75]    (10.93,-3.29) .. controls (6.95,-1.4) and (3.31,-0.3) .. (0,0) .. controls (3.31,0.3) and (6.95,1.4) .. (10.93,3.29)   ;
	\draw    (146.1,89.17) .. controls (191.55,114.33) and (199.02,114.1) .. (244,88) ;
	\draw [shift={(144,88)}, rotate = 29.09] [color={rgb, 255:red, 0; green, 0; blue, 0 }  ][line width=0.75]    (10.93,-3.29) .. controls (6.95,-1.4) and (3.31,-0.3) .. (0,0) .. controls (3.31,0.3) and (6.95,1.4) .. (10.93,3.29)   ;
	\draw    (263.1,89.17) .. controls (308.55,114.33) and (316.02,114.1) .. (361,88) ;
	\draw [shift={(261,88)}, rotate = 29.09] [color={rgb, 255:red, 0; green, 0; blue, 0 }  ][line width=0.75]    (10.93,-3.29) .. controls (6.95,-1.4) and (3.31,-0.3) .. (0,0) .. controls (3.31,0.3) and (6.95,1.4) .. (10.93,3.29)   ;
	\draw    (261,82) .. controls (306.87,53.78) and (314.19,53.5) .. (359.61,81.15) ;
	\draw [shift={(361,82)}, rotate = 211.29] [color={rgb, 255:red, 0; green, 0; blue, 0 }  ][line width=0.75]    (10.93,-3.29) .. controls (6.95,-1.4) and (3.31,-0.3) .. (0,0) .. controls (3.31,0.3) and (6.95,1.4) .. (10.93,3.29)   ;
	\draw    (382,82) .. controls (427.87,53.78) and (435.19,53.5) .. (480.61,81.15) ;
	\draw [shift={(482,82)}, rotate = 211.29] [color={rgb, 255:red, 0; green, 0; blue, 0 }  ][line width=0.75]    (10.93,-3.29) .. controls (6.95,-1.4) and (3.31,-0.3) .. (0,0) .. controls (3.31,0.3) and (6.95,1.4) .. (10.93,3.29)   ;
	\draw    (383.1,89.17) .. controls (428.55,114.33) and (436.02,114.1) .. (481,88) ;
	\draw [shift={(381,88)}, rotate = 29.09] [color={rgb, 255:red, 0; green, 0; blue, 0 }  ][line width=0.75]    (10.93,-3.29) .. controls (6.95,-1.4) and (3.31,-0.3) .. (0,0) .. controls (3.31,0.3) and (6.95,1.4) .. (10.93,3.29)   ;
	
	\draw (128,78) node [anchor=north west][inner sep=0.75pt]   [align=left] {0};
	\draw (247,78) node [anchor=north west][inner sep=0.75pt]   [align=left] {2};
	\draw (364,78) node [anchor=north west][inner sep=0.75pt]   [align=left] {4};
	\draw (490,78) node [anchor=north west][inner sep=0.75pt]   [align=left] {6};
	\draw (72,78) node [anchor=north west][inner sep=0.75pt]   [align=left] {Q:};
	\draw (188,38) node [anchor=north west][inner sep=0.75pt]   [align=left] {$a_*$};
	\draw (305,38) node [anchor=north west][inner sep=0.75pt]   [align=left] {$b_*$};
	\draw (426,38) node [anchor=north west][inner sep=0.75pt]   [align=left] {$c_*$};
	\draw (188,112) node [anchor=north west][inner sep=0.75pt]   [align=left] {$a^*$};
	\draw (305,112) node [anchor=north west][inner sep=0.75pt]   [align=left] {$b^*$};
	\draw (426,112) node [anchor=north west][inner sep=0.75pt]   [align=left] {$c^*$};

\end{tikzpicture}
\end{center}
We could label the vector spaces of these gauge bundles on different holomorphic varieties as $V_i$ with $i\in (0,2,4,6)$ in each vertex $Q_0$. The vertex operators would be $\mathcal{V}_i^{\vee}\otimes \mathcal{V}_j\otimes\wedge^m\mathcal{N}_{S_j/X},$ where $\mathcal{V}_i\in V_i$, $\mathcal{V}_j\in V_j$ and $S_j$ is the consistent algebraic varieties. Therefore, we can get the relating $Ext$ groups as
\begin{equation}
	Ext^{n}_{X}(\nu^{*}\mathcal{V}_i, \mu_{*}\mathcal{V}_j) \cong \sum_{p+q=n} H^{p}(S, \mathcal{V}_i^{\vee} \otimes \mathcal{V}_j \otimes \wedge^{q} \mathcal{N}_{S_j/X}),
\end{equation}
where $\nu_*$ and $\mu^*$ are the compositions of $(a_*,b_*,c_*)$ and $(a^*,b^*,c^*)$ respectively. In the case of $i=j$, the quiver vertices give a trivial path $e_i$ and the $Ext$ groups count BRST boundary states on homogeneous varieties.

\subsection{D2-branes wrap on $\mathbb{P}_1$}
Consider a smooth rational curve $C \cong \mathbb{P}_1$ fiberizing on a Calabi-Yau threefold. The Normal bundle $\mathcal{N}_{C/X}$ of $C$ in $X$ is $\mathcal{O}(a) \oplus \mathcal{O}(b)$.

If $\mathcal{N}_{C/X} \cong \mathcal{O}(-1) \oplus \mathcal{O}(-1)$, $C$ is infinitesimally rigid and $H^0(\mathcal{N}_{C/X})=0$ \cite{Katz:1986vil,1994Rational}. This leads to Clemens' conjecture\renewcommand{\thefootnote}{\roman{footnote}}\footnote{{\bf Clemens' conjecture:} \emph{A general quintic threefold contains only finitely many rational curves of degree d, for any $d \in \mathbb{Z}$. These curves are all infinitesimally rigid}.} \cite{1984Topics}. For example, we are about to use the Fermat quintic threefold $x_0^5+x_1^5+x_2^5+x_3^5+x_4^5+x_5^5 = 0$ that contains the family of lines in the homogeneous coordinates $(\mu,\nu)$ of $\mathbb{P}_1$ by $(\mu,-\mu,a\nu,b\nu,c\nu)$ with $a^5+b^5+c^5=0$. This could be described by a superpotential in the world-volume theory: $$W=\phi\psi^2.$$ where $\phi$ are complex structure moduli and $\psi$ are curve moduli.

If $\mathcal{N}_{C/X} \cong \mathcal{O} \oplus \mathcal{O}(-2)$, $C$ deforms to first order and then $dimH^0(\mathcal{N}_{C/X})=1$. The deformation is obstructed at $n$'th order and the superpotential description is $$W=\psi^{n+1}.$$

If $\mathcal{N}_{C/X} \cong \mathcal{O}(1) \oplus \mathcal{O}(-3)$, $C$ has a 2 parameter space of infinitesimal deformations and then $dimH^0(\mathcal{N}_{C/X})=2$. The superpotential is $$W(\rho,\psi)=\rho^3\psi^3+\phi F(\rho,\psi)+\cdots,$$ where the $F(\rho,\psi)$ is the generalization of $\phi\psi^2$.

Boundary-boundary OPE's are described as $$Exp^{p}(\mathcal{E}, \mathcal{F}) \times Ext^{q}(
\mathcal{F}, \mathcal{G}) \longrightarrow Ext^{p+q}(\mathcal{E}, \mathcal{G}).$$

The Yoneda pairing for $\mathbb{P}_1$ is $$Ext^{1}(\mathcal{O}_{\mathbb{P}_{1}}, \mathcal{O}_{\mathbb{P}_{1}}) \times Ext^{1}(\mathcal{O}_{\mathbb{P}_{1}}, \mathcal{O}_{\mathbb{P}_{1}}) \longrightarrow Ext^{2}(\mathcal{O}_{\mathbb{P}_{1}}, \mathcal{O}_{\mathbb{P}_{1}})$$. 

For the normal bundle $\mathcal{N}_{C/X} \cong \mathcal{O} \oplus \mathcal{O}(-2)$, both $Ext^{1}$ and $Ext^{2}$ above are one-dimensional. In fact, $$Ext^{1}(\mathcal{O}_{\mathbb{P}_{1}}, \mathcal{O}_{\mathbb{P}_{1}}) = H^{0}(\mathcal{N}_{C/X}) = \mathbf{C}$$

$$Ext^{2}(\mathcal{O}_{\mathbb{P}_{1}}, \mathcal{O}_{\mathbb{P}_{1}}) = H^{1}(\mathcal{N}_{C/X}) = \mathbf{C}.$$

For the normal bundle $\mathcal{N}_{C/X} \cong \mathcal{O}(1) \oplus \mathcal{O}(-3)$, both $Ext^{1}$ and $Ext^{2}$ above are two-dimensional. In fact, $$Ext^{1}(\mathcal{O}_{\mathbb{P}_{1}}, \mathcal{O}_{\mathbb{P}_{1}}) = H^{0}(\mathcal{N}_{C/X}) = \mathbf{C}^2$$

$$Ext^{2}(\mathcal{O}_{\mathbb{P}_{1}}, \mathcal{O}_{\mathbb{P}_{1}}) = H^{1}(\mathcal{N}_{C/X}) = \mathbf{C}^2.$$

Only the constant sheaf $\mathcal{O}$ has the holomorphic section. In the BCFT, the vertex operator $\theta$ are described by $Ext$ groups, and $Ext^2$ groups describe the vertex operator $\eta\theta$ (i.e. three fermion states) with Feynman diagram:

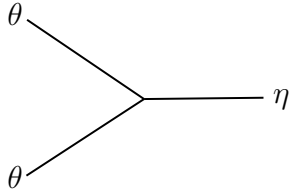
\tikzset{every picture/.style={line width=0.75pt}} 
\begin{center}
\begin{tikzpicture}[x=0.75pt,y=0.75pt,yscale=-1,xscale=1]
	
	\draw    (101.33,141.22) -- (160.33,181.22) ;
	\draw    (101,219.89) -- (160.33,181.22) ;
	\draw    (220.56,180.56) -- (160.33,181.22) ;
	
	\draw (90,131) node [anchor=north west][inner sep=0.75pt]   [align=left] {$\theta$};
	\draw (90,213.67) node [anchor=north west][inner sep=0.75pt]   [align=left] {$\theta$};
	\draw (224,175) node [anchor=north west][inner sep=0.75pt]   [align=left] {$\eta$};

\end{tikzpicture}
\end{center}

This diagram means that the BCFT correlation function are encoded by three copies of Yoneda pairing, giving two $\theta$'s and generating a $\eta\theta$. Therefore, the result is that a correlation function involves one $\eta$ and two $\theta$'s.One available BCFT interation term would be $$\int_{\partial\Sigma}F_{i\overline{j}}\rho^{i}\eta^{\overline{j}}.$$ The $\rho$ -- $\theta$ contraction would generate a propagator factor proportional to $1/z$, and the boundary integral would be scale-invariant. We assume that the curvature of the Chan-Paton factors is trivial in all present case.

\subsection{D4-branes wrap on K3 fiberation}
We consider the two parameter models which are the K3 fibered Calabi-Yau treefolds $\mathcal{M}$, obtained by resolving singularities of degree eight hypersurfaces $\widehat{M} \subset  \mathbb{P}(1,1,2,2,2)$ or $\mathbb{P}(1,1,2,2,6) $. The polynomials for these two hypersurfaces are \cite{1993Mirror}$$ p = x^{8}_{1} + x^{8}_{2}  + x^{4}_{3}  + x^{4}_{4}  + x^{4}_{5}$$  $$ p = x^{12}_{1} + x^{12}_{2}  + x^{6}_{3}  + x^{6}_{4}  + x^{2}_{5}.$$

We firstly consider the $\widehat{M} \subset \mathbb{P}(1,1,2,2,2)$, we can get that$$x^{8}_{1} + x^{8}_{2}  + x^{4}_{3}  + x^{4}_{4}  + x^{4}_{5} = 0  \stackrel{x_{2} = \lambda x_{1}}{\Longrightarrow} (1 +  \lambda^{8})x^{8}_{1} + x^{4}_{3} + x^{4}_{4} + x^{4}_{5} = 0 $$ $$\stackrel{y_{1} = x^{2}_{1}}{\Longrightarrow} (1 + \lambda^{8})y^{4}_{1} + x^{4}_{3} + x^{4}_{4} + x^{4}_{5} = 0.$$ This equation is the linear system $|L|$ of divisors and it is a pencil of quatic K3 surface.  There is a curve $C$ of singularities which is described by $$x_1=x_2=0, \qquad x_3^4+x_4^4+x_5^4=0.$$ We could get the exceptional divisor $E$ by blowing up the curve $C$ on which every point translates into a $\mathbb{P}_1$. The relation between $L$ and $E$ can be represent as $$|H| = |2L + E|.$$. 

Since $L\cdot L=0$, we have $$H\cdot L^2 = 0, \qquad L^3 = 0.$$ Let $l = \frac{1}{2}H \cdot E, \quad  h = \frac{1}{2} H\cdot L$, We have the intersection relations between systems and curves: $$ L \cdot l = 2, \qquad L\cdot h = 0;$$ $$H \cdot l = 0 \qquad H \cdot h = 2.$$

D4-branes wrap on K3 fiberation space $S$ and we can get the normal bundle of $S$ in $\mathcal{M}$: $$\mathcal{N}_{S/M} \cong \mathcal{O}(S)|_{S}.$$ So the $Ext$ groups of gauge bundles on D4-branes could be$$Ext^{n}_{X}(i_{*}\mathcal{E}, i_{*}\mathcal{E}) \cong \sum_{p+q=n} H^{p}(S, \mathcal{E}^{\vee} \otimes \mathcal{E} \otimes \wedge^{q}\mathcal{N}_{S/X}) = \sum_{p+q=n} H^{p}(\mathcal{O}^{q}(S)).$$
The multipoint (i.e. 4pt) correlator is described by the Conformal Partial Waves (CPW) functions \cite{2018epfl}. CPW need five copies of Yoneda pair to describe the correlation functions: four $\theta$'s and one $\eta$ generating two $\theta\eta$'s with Feynman diagram as

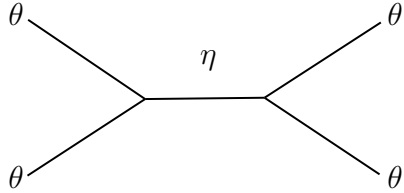
\tikzset{every picture/.style={line width=0.75pt}} 
\begin{center}
\begin{tikzpicture}[x=0.75pt,y=0.75pt,yscale=-1,xscale=1]
	
	\draw    (101.33,141.22) -- (160.33,181.22) ;
	\draw    (101,219.89) -- (160.33,181.22) ;
	\draw    (220.56,180.56) -- (160.33,181.22) ;
	\draw    (220.56,180.56) -- (278.56,219.22) ;
	\draw    (220.56,180.56) -- (279.22,142.56) ;
	
	\draw (90,131) node [anchor=north west][inner sep=0.75pt]   [align=left] {$\theta$};
	\draw (90,213.67) node [anchor=north west][inner sep=0.75pt]   [align=left] {$\theta$};
	\draw (186.67,155.33) node [anchor=north west][inner sep=0.75pt]   [align=left] {$\eta$};
	\draw (281,131) node [anchor=north west][inner sep=0.75pt]   [align=left] {$\theta$};
	\draw (281,213.67) node [anchor=north west][inner sep=0.75pt]   [align=left] {$\theta$};

\end{tikzpicture}
\end{center} 
The available interaction term would be the boundary interactions $$\int_{\partial\Sigma}F_{i_1\overline{j}i_2}\rho^{i_1}\eta^{\overline{j}}\rho^{i_2}$$. We could also contract the $\rho$ on one of the $\theta$'s, leaving us with two $\theta$'s and one $\eta$, perfect to match the available zero modes. The $\rho - \theta$ contraction would generate a propagator factor proportional to $1/z$ for $\int_{\partial\Sigma}F_{i\overline{j}}\rho^{i}\eta^{\overline{j}}$, two propagator factors proportional to $1/{z_{1}}\cdot1/{z_{2}}$ for $\int_{\partial\Sigma}F_{i_1\overline{j}i_2}\rho^{i_1}\eta^{\overline{j}}\rho^{i_2}$, and the boundary integral would give a scale-invariant result.

\subsection{D4-D2 branes boundary interaction}
We have declared that it is impossible to count all the D4-D2 gauge bundles interaction cohomology classes, because we can not classify all these gauge bundles on generic points. We will study the Gepner model in next section to get more valuable information. However, from the cohomology groups in ($\ref{7}$) and ($\ref{8}$), we can obvious that the most fundamental terms are the two normal gauge bundles $\mathcal{N}_{S/X}$ and $\mathcal{N}_{Y/X}$. Therefore, from our analysis of these $Ext$ groups and in order to embrace on the more easily and more meaningfully physical interpretation, we may analyze the cup product of ($\ref{7}$) and ($\ref{8}$) as$$Ext^{m}_{X}(i_{*}\mathcal{E}, \pi^{*}\xi) \times Ext^{n}_{X}(\pi^{*}\xi, i_{*}\mathcal{E}).$$ This product is not isomorphism with $$ Ext^{m+n}_{X}(i_{*}\mathcal{E}, i_{*}\mathcal{E}),$$ since these $Ext$ groups do not contain the normal gauge bundles of $Y$. We need to start calculating with the cohomology groups:
$$Ext^{m}_{X}(i_{*}\mathcal{E}, \pi^{*}\xi) \times Ext^{n}_{X}(\pi^{*}\xi, i_{*}\mathcal{E}) \longrightarrow \sum_{p+q=m}\sum_{k+l=n}H^{p}(Y, \wedge^{q}\mathcal{N}_{Y/X}) \otimes H^{k}(S, \wedge^{l}\mathcal{N}_{S/X})$$
This cup product gives the cohomology groups: $$H^{*}(Y \times S, \wedge^{**}\mathcal{N}_{Y\times S/X}).$$

The available interaction terms described by CPW and OPE consistent with these $Ext$ groups should be
$$\int_{\partial\Sigma_{Y}}F_{i\overline{j}}\rho^{i}\eta^{\overline{j}}\int_{\partial\Sigma_{S}}F_{i_1\overline{j}i_2}\rho^{i_1}\eta^{\overline{j}}\rho^{i_2}.$$
The product of these two boundary integrals would be scalar invariant. 
\section{Sheaf data on D4-D2-branes system}

We consider the prepotential of the two parameter models where we mainly focus on the Calabi-Yau treefolds $M$ obtained by $\mathbb{P}^{(1,1,2,2,2)}[8]$. The $N=2$ special geometry is encoded in a prepotential whose classical part can be written in the form \cite{1993Mirror}:$$\mathcal{F}(t_{1}, t_{2}) = -\frac{1}{3!}c_{111}t^{3} -\frac{1}{2}c_{112}t^{2}_{1}t_{2} + \frac{1}{24}(b_{1}t_{1} + b_{2}t_{2}) + const,$$ where $t_{1}$ and $t_{2}$ paprametrize the complexified K{\"a}hler cone: $K = t_{1}J_{1} + t_{2}J_{2}$. $J_{1}=H$ and $J_{2}=L$ are the K{\"a}hler classes of the K3 fiberation and $\mathbb{P}_{1}$ base respectively.

The topological intersection coefficients for the two parameter models are \cite{1993Mirror}
$$\mathbb{P}^{(1,1,2,2,2)}[8]:\qquad c_{111}=8,\quad c_{112}=4,\quad b_{1}=56,\quad b_{2}=24$$
$$\mathbb{P}^{(1,1,2,2,6)}[12]:\qquad c_{111}=4,\quad c_{112}=2,\quad b_{1}=52,\quad b_{2}=24$$

We have the symplectic period vector $\Pi = (2\mathcal{F} - t_{i}\partial_{i}\mathcal{F},\mathcal{F}_{1},\mathcal{F}_{2},1,t_{1},t_{2})$, reading for $\mathbb{P}^{(1,1,2,2,2)}[8]$:
\begin{equation}
	\left(
	\begin{array}{c}
		\frac{4}{3}t_1^3+2t_1^2t_2+\frac{7}{3}t_1+t_2\\
		-4t_1^2-4t_1t_2+\frac{7}{3}\\
		-2t_1^2+1\\
		1\\
		t_{1}\\
		t_{2}\\
	\end{array}
	\right)
\end{equation}

The central charge of a D-brane configuration is:
\begin{equation}\label{10}
	Z(Q(V)) = -\int e^{-K}\wedge Q = -\int e^{-K}\wedge Tr(e^{F})\wedge\sqrt{\hat{A}(X)},
\end{equation}
where $Q = (r, c_{1}(V), ch_2(V) + \frac{r}{24}c_{2}(X), ch_{3}(V) + \frac{1}{24}c_{1}(V)c_{2}(X)) \in H^{2*}(X)$ is the Mukai vector of effective brane charges, $r$ is the rank of the bundle and $ch_i(V)$ are the Chern characters of the Chan-Paton sheaf $V$. From this equation, we can get a topological invariant of $K$-theory:
\begin{equation}\label{11}
	Z(n_i) = n_{6}\mathcal{F}_0 + n_{4}^{1}\mathcal{F}_1 + n_{4}^{2}\mathcal{F}_{2} + n_{0} + n_{2}^{1}t_{1} + n_{2}^{2}t_{2}.
\end{equation}                                                                                                       From these two equations, we have
\begin{equation}
	\left(
	\begin{array}{c}
		r = n_{6}\\ 
		c_{1}(V) = \frac{1}{2}n_4^1J_1 + (n_4^2 - n_4^1)J_2\\
		ch_2(V) = n_2^1h + n_2^2l\\
		ch_3(V) = -n_0 - \frac{1}{3}n_4^1 - 2n_4^2
	\end{array}
	\right)
\end{equation}                                                      

For our D4-D2-branes interaction system, we should firstly set $r=0$, $n_4^1$ cannot be equal to zero together and $n_2^2 \neq 0$. For the pure D4-branes, the $Q$ vector reads \cite{2000D} $$Q = (0, n_4^2J_2, n_2^1h, -n_0 - 2n_4^2).$$
For the D2-branes wrapping on $\mathbb{P}_1$, $Q$ reads $$Q = (0, 0 ,n_2^2l, -n_0)$$

\subsubsection{The Mukai vector}
For D6-branes wrapping on the whole Calabi-Yau manifold, the Mukai vector is represented as \cite{2001D}
$$
Q=\left(r, c_1(V), \operatorname{ch}_2(V)+\frac{r}{24} c_2(X), \operatorname{ch}_3(V)+\frac{1}{24} c_1(V) c_2(X)\right),
$$ where $V$ is stable \cite{1998On} for conserving the D-brane configuration supersymmetry. The central charge is $$
Z(Q)=\frac{r}{6} t^3-\frac{1}{2} \operatorname{ch}_1(V) t^2+\left(\operatorname{ch}_2(V)+\frac{r}{24} c_2(X)\right) t-\left(\operatorname{ch}_3(V)+\frac{1}{24} c_1(V) c_2(X)\right)
$$

For D4-branes wrapping on the K3 fiberation and the including $i: S \hookrightarrow X$. The Mukai vector can be computed by a simple application of the Grothendieck-Riemann-Roch formula for this embedding $i$ \cite{0The}.$$i_*(ch(V)Td(D)) = ch(i_*V)Td(X),$$ where $V$ is the coherent sheaf on the fiberation. The form of Mukai vector for D4-branes is \cite{2001D}
$$Q=(0,rD,i_*c_1(V)+\frac{r}{2}i_*c_1(D),ch_2(V)+\frac{1}{2}c_1(V)c_1(D) +\frac{r}{8}c_1{D}^2+\frac{r}{24}c_2(D))$$
The central charge in the large volume limit is
$$
\begin{aligned}
	Z(Q)=&-\frac{r}{2} t^2 D+\left(i_* c_1(V)+\frac{r}{2} i_* c_1(D)\right) t-\operatorname{ch}_2(V)-\\
	& \frac{1}{2} c_1(V) c_1(D)-\frac{r}{8} c_1(D)^2-\frac{r}{24} c_2(D)
\end{aligned}
$$
\subsection{Gepner model}
A Gepner model is a rational 2D SCFT which is a tensor product of $N=2$ super-minimal model CFT. This means that Gepner models appear as the limiting cases of sigma-models with target space a $6d$ Calabi-Yau manifold at singular points in the moduli space of the CY target rather than as sigma-models with target manifold a smooth manifold \cite{1989gepner,1990gepner}. The $\mathcal{N} = 2$ minimal models at level $k$ are SCFTs with central charge $c = \frac{3k}{k+2}< 3$. The primary fields $\Phi^l_{m,s}$ are labeled by three integers $l, m, s$ with$$l= 0, 1, \cdots ,k, \quad m=-(k+1), \cdots, k+2, \quad s=0,2,\pm 1.$$ where $l$ and $m$ label the $SU(2)$ gauge filds and $s=0$ and $s=\pm1$ respectively determine the boundary condition for the $NS$ sector and the other one for the $R$ sector. The conform dimension $h$ and $U(1)$ charge $q$ of the primary fields are given by$$h = \frac{l(l+2)-m^2}{4(k+2)}+\frac{s^2}{8},$$ $$q=\frac{m}{k+2} - \frac{s}{2}$$.

The boundary states can be labeled by Cardy's notation by $\alpha = (L_j,M_j,S_j)$. Let $K=lcm\{h_1, \cdots, h_i\}$, where $h_i=k_i+2$ are the heights of minimal models. The weights of minimal model would be $\omega_j=K/h_j$. In order to get the geometric interptretation of the boundary states, it is conventional to calculate the intersection of these branes. This observable for B-type boundary states is: 
\begin{equation}\label{13}
 I_B = \frac{1}{C}(-1)^{\frac{S-\tilde{S}}{2}}\sum_{m_j'}\delta_{\frac{M-\tilde{M}}{2}+\sum\frac{K'}{2k_j+4}(m_j'+1)}\prod_{j=1}^{r}N_{L_j,\tilde{L}_j}^{m_j'-1}.
\end{equation}
As explained before, the Gepner point is a special point in the moduli space where the superconformal field theory is exactly solvable. This nature allows us to obtain valuable information on the gauge bundles (coherent sheaves) and the spectrum of BPS states. For example, The construction of B-type boundary state in the $(k=3)^5$ continuously connects to the coherent sheaves construction on the quintic, generated by the $Z_5^3$ symmetry \cite{Philip1991A}. 

For the two parameters models, the quantum mirror of $\mathbb{P}^{(1,1,2,2,2)}[8]$ satisfies a $Z_8$ action, whose continuously connecting Gepner model could be represented as $(k=6)^2(k=2)^3$. The natural basis of quantum $Z_8$ discrete symmetry could be described by a eight-vector $(\omega_0,\cdots,\omega_7)^t$. These eight periods satisfy the relations: \cite{1993Mirror} $$\omega_0+\omega_2+\omega_4+\omega_6=0$$$$\omega_1+\omega_3+\omega_5+\omega_7=0.$$ Therefore, we have six independent periods $(\omega_0,\cdots,\omega_5)^t$. Gepner model for $\mathbb{P}^{(1,1,2,2,6)}[12]$ would be $(k=10)^2(k=4)^3$. The boundary states for two parameter models could be expressed as $$ |L_1,L_2,L_3,L_4,L_5,M,S \rangle,$$ where $M\in Z_{16}$ for $\mathbb{P}^{(1,1,2,2,2)}[8]$ and $M\in Z_{24}$ for $\mathbb{P}^{(1,1,2,2,2)}[12]$. Each fusion coefficient $N_{L_j,\tilde{L}_{j}}$ in the intersection matrix (\ref{13}) for the two parameter models could be expressed by a matrix $g$\renewcommand{\thefootnote}{\roman{footnote}}\footnote{$g^{\frac{1}{2}}=r$ is the shift matrix.} \cite{2001D}:
$$
n_{L, \tilde{L}}=n_{\tilde{L}, L}=g^{\frac{|L-\tilde{L}|}{2}}+g^{\frac{|L-\tilde{L}|}{2}+1}+\cdots+g^{\frac{L+\tilde{L}}{2}}-g^{-1-\frac{|L-\tilde{L}|}{2}}-\cdots-g^{-1-\frac{L+\tilde{L}}{2}}.
$$
The intersection for the ground state $|0,0,0,0,0\rangle$ would be $$I_B=\prod_{j}(1-g_j^{-1})=(1-g^{-1})^2(1-g^{-2})^3=9-9g,$$ where $g$ is $g_j=g^{\omega_j}$.

In the fixed orbits, the higher boundary state excited operators for the intersection matrix could be defined as
$$t_{L_j} = t_{L_j}^{t} = \sum_{l=-\frac{L_j}{2}}^{\frac{L_j}{2}}g^{l}_j.$$ Each factors $N_{L_j,\tilde{L}_j}^{m_j'-1}$ in the intersection matrix (\ref{13}) could be excited by $t_L, t_L^t$. The relations of $n_{0,0}$ ,$n_{L,0}$ and $n_{L,L}$ are $$t_{L_j}n_{0,0} = n_{L_j,0}, \qquad n_{L_j,0}t_{\tilde{L}_j} = n_{L_j,\tilde{L}_j}.$$ 
The charge of the boundary state $q_{B}$ in the Gepner basis is given by $$q_G = q_Bt_{L_1} t_{L_2}t_{L_3}t_{L_4}t_{L_5}(1-g_j)$$.

\subsubsection{Marginal operators}

The number of marginal operators related to the Witten index in the NS sector which give a proof of Mukai's formula \cite{1984Symplectic,1987Moduli} for the dimension of the moduli space of $1/2$-BPS states. The number of marginal operators is counted by the absolute values of the fusion matrices: $$\frac{1}{2}\prod_{j}|n_{L_iL_j}(|g|)|-vac,$$ where $vac$ are the vacuum values of boundary states. We count the number of boundary marginal operators for one single states $|L_j,M,S\rangle$ and split states $|(L_j,0)(0,0,0)\rangle$ in different $\omega_j$ as\\
\begin{center}
\begin{tabular}{lcc}
	\toprule 
	$|L_1,\quad L_2,\quad L_3,\quad L_4,\quad L_5\rangle$ & marginal operator  &  vacuum \\ 
	\midrule 
	$|0,$\qquad$0\rangle$ & 0 & 1 \\
    $|1,$\qquad$0\rangle$ & 0 & 1 \\
	$|2,$\qquad$0\rangle$ & 2 & 1 \\
	$|3,$\qquad$0\rangle$ & 3 & 1 \\
	$|4,$\qquad$0\rangle$ & 4 & 1 \\
	$|5,$\qquad$0\rangle$ & 6 & 1 \\
	$|6,$\qquad$0\rangle$ & 6 & 2 \\
	\qquad \qquad\qquad$|0,\quad0,\quad0\rangle$ & 0 & 1 \\
	\qquad \qquad\qquad$|1,\quad0,\quad0\rangle$ & 0 & 1 \\
	\qquad \qquad\qquad$|2,\quad0,\quad0\rangle$ & 2 & 1 \\
	\qquad \qquad\qquad$|3,\quad0,\quad0\rangle$ & 3 & 1 \\
	\qquad \qquad\qquad$|4,\quad0,\quad0\rangle$ & 3 & 2 \\
	$|3,\quad0,\quad0,\quad0,\quad0,\quad0\rangle$ & 3 & 1 \\
	$|6,\quad0,\quad0,\quad0,\quad0,\quad0\rangle$ & 6 & 2 \\
	\bottomrule 
\end{tabular}
\end{center}

It is obvious that if the two boundary states are the same, one vacuum introduces one spectral flow operator in the open string channel. If they are not the same, neither state propagates and the brane could be viewed as a single object only satisfying the $U(1)$ symmetry. For our purpose, the D4-branes wrapped on $K3$ fiberation space whose compactifications are geometric throughout their moduli space \cite{1988Observations}. Coherent semistable sheaves could describe these compactifications. The Mukai vector $Q=v(V)$ labels the number of moduli space of these sheaves $V$.

\subsubsection{Sheaf's data on Gepner points}
We can get the analytical continuation conditions between the large-radius periods and the Gepner point:
$$Z=\vec{n}\cdot\vec{\Pi}=(\vec{n}^Gm^{-1})\cdot(m\vec{\Pi}^G),$$
where $m$ is the monodromy matrix and $\vec{n}$ is the coefficient vector of $Z(n_i)$ in (\ref{11}). The matrix $m$ depend on the $Sp(6,\mathbb{Z})$ monodromy transformation. For $\mathbb{P}^{(1,1,2,2,2)}[8]$, $m$ would be \cite{1993Mirror,2000D}
\begin{equation}
\left(
\begin{array}{cccccc}
	-1 & 1 & 0 & 0 & 0 & 0\\
	2 & 2 & \frac{1}{2} & \frac{1}{2}& \frac{1}{2}& \frac{1}{2}\\
	1 & 0 & 1 & 0 & 0 & 0\\
	1 & 0 & 0 & 0 & 0 & 0\\
	-\frac{1}{2} & 0 & \frac{1}{4} & 0 & -\frac{1}{4} & 0\\
	\frac{1}{2} &\frac{1}{2} & -\frac{1}{4} & \frac{1}{4}& \frac{1}{4} & -\frac{1}{4}\\
\end{array}
\right)
\end{equation}
For $\mathbb{P}^{(1,1,2,2,6)}[12]$, $m$ would be 
\begin{equation}
	\left(
	\begin{array}{cccccc}
		-1 & 1 & 0 & 0 & 0 & 0\\
		\frac{3}{2} & \frac{3}{2} & \frac{1}{2} & \frac{1}{2} & -\frac{1}{2} & -\frac{1}{2}\\
		1 & 0 & 1 & 0 & 0 & 0\\
		1 & 0 & 0 & 0 & 0 & 0\\
		-\frac{1}{2} & 0 & \frac{1}{2} & 0 & \frac{1}{2} & 0\\
		\frac{1}{2} & \frac{1}{2} & -\frac{1}{2} & \frac{1}{2} & -\frac{1}{2} & \frac{1}{2}\\
	\end{array}
	\right)
\end{equation}
The charge of the boundary state $q_B$ in the Gepner basis is given by  $$q_G = q_Bt_{L_1} t_{L_2}t_{L_3}t_{L_4}t_{L_5}(1-g_j),$$ where $q_B$ is nonzero in the Mth state. The large-radius charges could be expressed as $q_L=m^{-1}q_G$. we list the data of several states.

\subsection{$L=|1,0,0,0,0 \rangle $}
\begin{center}
	\begin{tabular}{ccccccc}
		\toprule 
		$M$ \quad &$n_6$ \quad & \quad$n_4^1$\quad  &  \quad$n_4^2$ \quad& $n_0$ \quad& \quad$n_2^1$\quad & \quad$n_2^2$ \\ 
		\midrule 
	$m_1$ & 0	& 1 & 0 & -2 & 0 & 1\\
	$m_2$ &	0 & 1 & 0 & -1 & 0 & 0\\
	$m_3$ &	1 & -1 & -1 & 1 & 0 & 0 \\
	$m_4$ &	1 & 0 & 0 & 1 & -1 & -1 \\
	$m_5$ &		0	& 1 & 0 & 1 & 0 & -2\\
	$m_6$ &		1& -1 & -1 & -2 & 2 & 1\\
	$m_7$ &		2	& 1 & 1 & -1 & -3 & 0\\
	$m_8$ &		2	& 1 & -1 & -2 & -1 & 1\\
		\bottomrule 
	\end{tabular}
\end{center}
There are 8 charge vectors. The states $m_1$, $m_5$ are exactly the D4-D2 boundary states that we really wanted. The vector $m_1$ include a single D4-brane with interaction between sheaves on $K3$ fiberation and sheaves on $\mathbb{P}_1$ fiberation, a single D2-brane on $\mathbb{P}_1$ fiberation and two D0-antibranes, which continuously connects to the $Ext$ groups (\ref{7}) and (\ref{8}). 
We have use the $Ext$ groups (\ref{7}) and (\ref{8}) to count the D4-D2 interaction terms. Therefore, these groups could parameterize this sheaf $V$ (i.e. the pullback of Chan-Paton bundles $\xi^{\vee} \otimes \mathcal{E}$ on $X$) as $$V \cong i^*(\xi^{\vee} \otimes \mathcal{E}),$$ and the $Ext$ groups (\ref{7}) could be expressed as $$Ext^{n}_{X}(\pi^{*}\xi, i_{*}\mathcal{E}) \cong \sum_{p+q=n} H^{p}(S, i_* V \otimes \wedge^{q} \mathcal{N}_{S/X}).$$
There is a map from the sheaves on $\mathbb{P}_1$ fiberation to the sheaves on $K3$-$\mathbb{P}_1$ interaction fiberation as $i^*\pi^*\xi \hookrightarrow V$. The "semistable" condition for the coherent sheaves is that any subsheaf sequence $0 \rightarrow \mathcal{F} \rightarrow \mathcal{V}$ satisfies the slope inequality \cite{1998On}:
$$
\mu(\mathcal{F}) \equiv \frac{\int \omega^{d-1} c_1(\mathcal{F})}{\operatorname{ch}_0(\mathcal{F})} \leq \frac{\int \omega^{d-1} c_1(\mathcal{V})}{\operatorname{ch}_0(\mathcal{V})} \equiv \mu(\mathcal{V}),
$$  
where $\omega$ is the K\"{a}hler form. In this paper we will not delve into the stability of D4-D2-brane system, and it is still an open problem for the D4-D2-brane bound states. In the previous section, we stated that the Makui vector determines the number of moduli spaces of these sheaves. In terms of the rank $r$ and Chern classes $c_i$ of $V$, the Mukai vector is represented as \cite{1998On,1999D} $$v(V)=(r, c_1, \frac{1}{2}c_1^2-c_2+r).$$ Mukai's theorem \cite{1984Symplectic,1987Moduli} states that the complex dimension of the moduli space of an irreducible coherent sheaf $V$ is $$Dim(\mathcal{M}(V))=\langle v(V), v(V)\rangle +2.$$ These states for the inner product $v^2\geqslant-2$ decompose into BPS and anti-BPS states. BPS states have $r>0$ or $r=0$, $c_1>0$, or $r=c_1=0,ch_2(V)>0$.

The explicit Mukai vector of the states $m_1$ would be $$ Q = v(V)=(0, \quad\frac{1}{2}E, \quad l,\quad \frac{8}{3}),$$ where $E$ is the exceptional divisor by blowing up singularities. It is obvious that the complex dimension of the moduli space of this coherent sheaf is greater than zero for $r=1$. There is a vacuum $v=1$, so this sheaf is irreducible.

This charge is gotten by twisting a nontrivial line bundle to the $K3$ fiberation by modifying the charge vector $m_2$ (i.e. $(0,\frac{1}{2}E,0,\frac{8}{3}))$ as we explained before. This D4-brane carry out an electrical charge and a magnetic charge with the Dirac quantization condition:
$$e\cdot g \in 2\pi \mathbb{Z}.$$ Through the magnetic flux, this state induces a lower charge on D2-brane wrapped on the $\mathbb{P}_1$ fiberation and two D0-antibrane charges. The states $m_5$ has the same D4-D2-branes system but the lower brane charges are replace by two D2-antibrane charges and a D0-brane charge.

We find two special D4-D2-brane interaction forms in the boundary state $|2,0,0,0,0\rangle$ \begin{center}
	\begin{tabular}{ccccccc}
		\toprule 
		$M$ \quad &$n_6$ \quad & \quad$n_4^1$\quad  &  \quad$n_4^2$ \quad& $n_0$ \quad& \quad$n_2^1$\quad & \quad$n_2^2$ \\ 
		\midrule 
		$m_1$ & 0	& 2 & -1 & 0 & 1 & -2\\
		$m_2$ &	0 & 1 & -2 & 0 & 2 & -1\\
		\bottomrule 
	\end{tabular}
\end{center}
The state vectors $m_1$ and $m_2$ are multiple branes. There is an open dense set consisting of the moduli space of holomorphic vector bundles for $r > 1$. The Mukai vector for the state $m_1$ is $Q=(0,3,h-2l,-\frac{5}{3})$. The holomorphic gauge bundles on the D4-D2-branes satisfy the $SU(2)$ gauge symmetry and there is a Higgs field $\Psi_i, i=1\cdots D$ for the vacuum $v=1$. The Higgs field for the $SU(2)$ gauge bundle satisfy the $U(1) \times U(1)$ symmetry:
$$
\Phi_i=\left(\begin{array}{cc}
	a_i^{(1)} \mathbf{1}_{1 \times 1} & 0 \\
	0 & a_i^{(2)} \mathbf{1}_{2 \times 2}.
\end{array}\right)
$$
Higgs field describes the normal motions of this D4-D2-multibranes.
However, this state do not satisfy the $SU(3)$ symmetry although $r=3$, but this symmetry is $SU(2)\times U(1)$. The system also include a single pure $K3$ fiberation D4-antibrane and a D2-brane wrapped on a curve in the $K3$ fiberation. The stability of these brane system need to be further studied, and then we will explore this issue in depth in our next work.
\bibliography{ref}{}

\begin{thebibliography}{10}

\bibitem{2003D}
A.~Caldararu, S.~Katz, and E.~Sharpe.
\newblock D-branes, {B} fields, and {E}xt groups.
\newblock {\em Advances in Theoretical \& Mathematical Physics}, 7(3):381--404,
  2003.

\bibitem{1998On}
J.~A. Harvey and G.~Moore.
\newblock On the algebras of {BPS} states.
\newblock {\em Communications in Mathematical Physics}, 197(3):489--519, 1998.

\bibitem{2002D}
S.~Katz and E.~Sharpe.
\newblock D-branes, open string vertex operators, and {E}xt groups.
\newblock {\em Advances in Theoretical \& Mathematical Physics},
  6(6):979--1030, 2002.

\bibitem{2000MIRROR}
David~A. Cox and Sheldon Katz.
\newblock Mirror symmetry and algebraic geometry.
\newblock {\em Mathematical {S}urveys and Moographs}, 64, 1999.

\bibitem{1999Anomalies}
D.~S. Freed and E.~Witten.
\newblock Anomalies in {S}tring {T}heory with {D}-branes.
\newblock {\em Asian Journal of Mathematics}, 3(4):819--851, 1999.

\bibitem{1988Space}
D.~Gepner.
\newblock Space-time supersymmetry in compactified string theory and
  superconformal models.
\newblock {\em Nuclear Physics B}, 296(4):757--778, 1988.

\bibitem{1993Mirror}
P.~Candelas, Xdl Ossa, A.~Font, S.~Katz, and D.~R. Morrison.
\newblock Mirror {S}ymmetry for {T}wo {P}arameter {M}odels -- {I}.
\newblock 1993.

\bibitem{alma9914090123903768}
A.~Kapustin, M.~Kreuzer, and K.-G.(Eds.) Schlesinger.
\newblock Homological mirror symmetry : new developments and perspectives,
  2009.

\bibitem{1999K3}
Bruce Hunt and Rolf Schimmrigk.
\newblock K3-fibered {C}alabi-{Y}au threefolds {I}, the twist map.
\newblock {\em International Journal of Mathematics}, 10(7), 1999.

\bibitem{1999D}
I.~Brunner, M.~R. Douglas, A.~Lawrence, and C.~R?Melsberger.
\newblock D-branes on the {Q}uintic.
\newblock {\em Journal of High Energy Physics}, 2000(8):339--358, 1999.

\bibitem{0Representation}
M.~Brion.
\newblock Representation of quivers.

\bibitem{0Representation1}
A.~Craw.
\newblock Explicit methods for derived categories of sheaves.
\newblock 2007.

\bibitem{0Representation2}
A.~Craw.
\newblock Quiver representations in toric geometry.
\newblock 2008.

\bibitem{Katz:1986vil}
S.~Katz.
\newblock {On the finiteness of rational curves on quintic threefolds}.
\newblock {\em Compos. Math.}, 60(2):151--162, 1986.

\bibitem{1994Rational}
S.~Katz.
\newblock Rational {C}urves on {C}alabi-{Y}au {T}hreefolds.
\newblock 1994.

\bibitem{1984Topics}
A.~Griffithsphillip and ZuckerSteven.
\newblock {\em {T}opics in {T}ranscendental {A}lgebraic {G}eometry. (AM-106)}.

\bibitem{2018epfl}
Slava Rychkov.
\newblock {EPFL} {L}ectures on {C}onformal {F}ield {T}heory in {D} $\ge$ 3
  {D}imensions.
\newblock 2018.

\bibitem{2000D}
P.~Kaste, W.~Lerche, C.~A. Lutken, and J.~Walcher.
\newblock D-{B}ranes on {K}3-{F}ibrations.
\newblock {\em Nuclear Physics B}, 582(1-3):203--215, 2000.

\bibitem{2001D}
D.~E. Diaconescu and C~Römelsberger.
\newblock D-{B}ranes and {B}undles on {E}lliptic {F}ibrations.
\newblock {\em Nuclear Physics B}, 574(1-2):245--262, 2000.

\bibitem{0The}
A.~Javanpeykar.
\newblock The {G}rothendieck-{R}iemann-{R}och theorem.

\bibitem{1989gepner}
D.~Gepner.
\newblock Space-{T}ime {S}upersymmetry in {C}ompactified {S}tring {T}heory and
  {S}uperconformal {M}odels - {S}cience{D}irect.
\newblock {\em Current Physics–Sources and Comments}, 4:381--402, 1989.

\bibitem{1990gepner}
A.~Gepner, D.~Strominger.
\newblock {L}ectures on {N}=2 string theory.
\newblock {\em Singapore: World Scientific}, 1990.

\bibitem{Philip1991A}
Philip Candelas, Xenia C. De~La Ossa, Paul~S. Green, and Linda Parkes.
\newblock A pair of {C}alabi-{Y}au manifolds as an exactly soluble
  superconformal theory.
\newblock {\em Nuclear Physics B}, 1991.

\bibitem{1984Symplectic}
S.~Mukai.
\newblock {S}ymplectic structure of the moduli space of sheaves on an abelian
  or k3 surface.
\newblock {\em Inventiones Mathematicae}, 77(1):101--116, 1984.

\bibitem{1987Moduli}
S.~Mukai.
\newblock Moduli of vector bundles on k3 surfaces and symplectic manifolds.
\newblock 1987.

\bibitem{1988Observations}
N.~Seiberg.
\newblock Observations on the moduli space of superconformal field theories.
\newblock {\em Nuclear Physics}, 303(2):286--304, 1988.

\end{thebibliography}
\bibliographystyle{unsrt}
\end{document}